\documentclass[aps,prb,showpacs,reprint,amsmath,amssymb,floats,superscriptaddress]{revtex4-1}
\usepackage{graphicx}
\usepackage{wasysym}

\usepackage{color}
\usepackage{bm}
\usepackage[caption=false]{subfig}
\usepackage{amsfonts}

\usepackage{appendix}

\begin{document}

\title{From kinetic to collective behavior in thermal transport on semiconductors and semiconductor nanostructures}

\author{C. de Tomas}
\affiliation{Department of Physics, Universitat Aut\`onoma de Barcelona, 08193 Bellaterra, Catalonia, Spain}

\author{A. Cantarero}
\affiliation{Materials Science Institute, University of Valencia, P. O. Box 22085, 46071 Valencia, Spain}

\author{A. F. Lopeandia}
\affiliation{Department of Physics, Universitat Aut\`onoma de Barcelona, 08193 Bellaterra, Catalonia, Spain}

\author{F. X. Alvarez}
\email{xavier.alvarez@uab.cat}
\affiliation{Department of Physics, Universitat Aut\`onoma de Barcelona, 08193 Bellaterra, Catalonia, Spain}

\date{\today}
\begin{abstract}
We present a model which deepens into the role that normal scattering has on the thermal conductivity in semiconductor bulk, micro and nanoscale samples. Thermal conductivity as a function of the temperature undergoes a smooth transition from a kinetic to a collective regime that depends on the importance of normal scattering events. We demonstrate that in this transition, the key point to fit experimental data is changing the way to perform the average on the scattering rates. We apply the model to bulk Si with different isotopic compositions obtaining an accurate fit. Then we calculate the thermal conductivity of Si thin films and nanowires by only introducing the effective size as additional parameter. The model provides a better prediction of the thermal conductivity behavior valid for all temperatures and sizes above 30 nm with a single expression. Avoiding the introduction of confinement or quantum effects, the model permits to establish the limit of classical theories in the study of the thermal conductivity in nanoscopic systems.
\end{abstract}



\maketitle

\section{Introduction}
\label{intro}
A general model, able to explain the thermal conductivity $\kappa$ in macro-, micro- and nanostructured systems is still an open challenge. Experimental measurements on low-dimensional semiconductors \cite{Asheghi1998,Li2003} have shown a drastic size-dependent reduction of $\kappa$ as compared to bulk values.\cite{Inyushkin2004,Glassbrenner1964} A great effort has been devoted in the last years to develop a general model which provides an accurate understanding of this reduction.\cite{Chantrenne2005,Mingo2003,Kazan2010,Balandin2001,Martin2009,Morelli2002} At present, we can confirm that, when the size of the samples is reduced, classical boundary effects are expected due to the existence of a surface or an interface. Below some nanometers phonon confinement may also influence thermal transport through the modification of the dispersion relations.

However, it is still under debate which are the most important effects at the different length scales since most of the proposed models do not agree even in the origin of the reduction of the thermal conductivity, whether it is due to a change in the relaxation times, a confinement or a quantum effect, especially within the range of $10-100$ nanometers. In order to obtain a thermal transport model valid at all ranges of sizes and temperatures, it is necessary to have some certainty about the limits of applicability of the classical approaches without the inclusion of the mentioned changes and the size dimensions where those are strictly necessary. 

Recent works have focused their attention on the calculation of phonon scattering rates by \textit{ab initio} techniques.\cite{Broido2007,Broido2010, Broido2012,Lindsay2013,Fugallo2013} These works suggest that the main reason for the poor adjustment of current theories arises from the use of empirical potentials with adjustable parameters or the use of classical expressions for the relaxation times. Based on \textit{ab initio} techniques, they solve numerically the Boltzmann transport equation (BTE) to obtain the scattering times and predict the thermal conductivity. In the last years, the thermal conductivity of several materials has been calculated (see Ref. \onlinecite{Broido2013} and references therein).  In these works, the theoretical predictions agree very well with the experiment in particular intervals of sizes and temperatures. Very recently, Fugallo \textit{et al.} \cite{Fugallo2013}, also using \textit{ab initio} techniques, calculate the thermal conductivity of bulk diamond and isotopically enriched diamond by solving the BTE using the variational principle and the conjugate gradient scheme. They introduced the scattering due to boundary effects with a shape factor to fit the low temperature region. The Mathiessen rule is used in both cases to account for the different scattering mechanisms.

In spite of these advances, the models based on \textit{ab initio} techniques looses some of the thermodynamics involved in the heat transport mechanisms, hidden behind the numerical complexity of the models. At this stage, a phenomenological model is always desirable when the physical processes can be clearly described. This perspective also aimed the work by Allen\cite{Allen2013}, where the widely used Callaway model\cite{Callaway1958} is improved by a more rigorous treatment of phonon-phonon scattering, paying particular attention to the introduction of the normal scattering relaxation time into the expression of the lattice thermal conductivity.

In the last decade, some authors have suggested that in order to predice nanoscale transport parameters, memory and nonlocalities had to be included \cite{Xavi2007,EIT1993} in the expressions. In this line of thought, here we demonstrate that some issues appearing when fitting thermal conductivity data are not related to the particular expression used for the relaxation times but with the way their thermodynamic averages are calculated. We show, from an approach based on Guyer and Krumhansl model, \cite{Krumhansl1965,Guyer1966,Guyer1966a} that a more appropriate equation for $\kappa$ can be obtained. This equation will provide a new insight into the underlying physics of thermal transport. It introduces a thermodynamic perspective that allows to understand the differences in phonon behavior in terms of the mixing rate of the different phonon-phonon processes. Our proposal is in good agreement with experimental data on bulk silicon,\cite{Inyushkin2004} thin films (TFs),\cite{Asheghi1998} and nanowires (NWs),\cite{Li2003} with characteristic sizes above 30 nm. We show that confinement or quantum effects are not necessary to understand the lattice thermal transport above these sizes and that the difficulty of prediction at the nanoscale seems to be deeply related to the thermodynamic treatment of phonon-phonon interactions. At the same time, this allows to establish a lower limit for classical models, where bulk properties are enough to understand the phenomenology. Only below this limit, of the order of a few tens of nanometers, confinement effects may play a role.\cite{Srivastava2007}

\section{Approaches to solve the Boltzmann transport equation}

BTE is the usual starting point in all thermal conductivity works. Its mathematical form and the physical interpretation of its terms in thermal transport applications have been widely discussed in the literature\cite{Peierls1955,Ziman1979}. Summarizing, when a small temperature difference $\delta T$ is applied on a system, the phonon distribution $f_{\bm{q}}$ moves from equilibrium at a linear rate. On the other hand, collisions turn the phonon distribution back to equilibrium at a rate that depends on the scattering transition rate. The BTE allows to obtain the resulting phonon distribution function by relating both rates
\begin{equation}
\label{boltz_0}
\left.\frac{\partial f_{\bm{q}}}{\partial t}\right|_{\textrm{\tiny{drift}}}= \left.\frac{\partial f_{\bm{q}}}{\partial t}\right|_{\textrm{\tiny{scatt}}}.
\end{equation}
Unfortunately, the analytical solution of Eq. (\ref{boltz_0}) is unknown. Two possible alternatives are: i) to solve the full equation numerically or ii) to derive some simplified expression replacing part of Eq. (\ref{boltz_0}) and solve it analytically.  

The computational power nowadays allows the numerical solution of the BTE in combination with density functional theory obtaining remarkable results in particular regions of temperature. Specifically, for silicon, comparison with natural and isotopically enriched bulk samples has been obtained in the [50-350] K range \cite{Ward2009}, but for low temperatures, the grid of $q$ points needed in this approach is out of the calculational capability \cite{Broido2012}. In order to improve this, Fugallo et al. \cite{Fugallo2013} combined \textit{ab initio} with a phenomenological expression with a fitting parameter for the boundary scattering for bulk diamond. In reduced size samples, predictions for wires have not been able to be compared with experiments since, in words of the authors, they provide larger values of $\kappa$ \cite{Broido2012}. To date this kind of approach has not been able to obtain a single solution valid at all ranges of temperatures and sample sizes for this material, demonstrating that phenomenological approaches are still necessary.

Here we develop an approach that allows to distinguish between two different regimes and shows that the difficulty for obtaining a global solution lies in the fact that each regime happens at different temperature intervals. In this section we define the terms that will be used in our model to obtain the final expression of the thermal conductivity.

In equilibrium it can be easily demonstrated that phonons follow the Bose-Einstein distribution function
\begin{equation}
\label{dist_equil_1}
f^{0}_{\bm{q}}=\frac{1}{e^{\hbar \omega_{\bm{q}}/k_{B}T}-1}\quad,
\end{equation}
where $\hbar\omega_{\bm{q}}\equiv\varepsilon_{\bm q}$ is the energy of the phonon mode ($\nu$,$\bm{q}$) (the branch $\nu$ will be omitted for simplicity), $T$ the absolute temperature and $k_{B}$ the Boltzmann constant.

If a temperature difference is applied on the system $\delta T$, an asymmetry in $f_{\bm{q}}$ will be generated in the direction of the resulting gradient $\nabla T$. In general, the final form of the distribution can be very complex, but under small $\delta T$ the deviations from equilibrium are expected to be small. In that case, we can expand $f_{\bm{q}}$ and keep the first term in the expansion:
\begin{equation}
\label{aprox_non_eq_dist}
f_{\bm{q}}\simeq f^{0}_{\bm{q}} + \frac{\partial f_{\bm{q}}}{\partial \varepsilon_{\bm{q}}}\delta \varepsilon\simeq
f^{0}_{\bm{q}} +  \frac{\partial f^{0}_{\bm{q}}}{\partial \varepsilon_{\bm{q}}}\Phi_{\bm{q}} = f^{0}_{\bm{q}} +  \frac{f^{0}_{\bm{q}}(f^{0}_{\bm{q}}+1)}{k_{B}T}\Phi_{\bm{q}}\quad ,
\end{equation}
where $\Phi_{\bm{q}}$ is a smooth function of the energy and temperature whose precise form depends on the scattering processes. Expressed in these terms, solving BTE is reduced to obtain an expression for $\Phi_{\bm{q}}$, which will lead to a thermal conductivity equation. The approach used to solve the problem will depend ultimately on the expected form of $\Phi_{\bm{q}}$.

There are roughly two main approaches to solve BTE analytically: the kinetic methods (KM) and the variational methods (VM). KM can be applied when the distribution function is expected to be very close to equilibrium. In this case, the collision term is usually simplified by assuming that it is proportional to the inverse of a relaxation time (relaxation time approximation (RTA)), depending only on the values of a single mode. Finding relaxation times for reduced regions of temperature and size is not difficult. The problem appears if one wants to extend the region of applicability to wider intervals using the same KM approach with the same RTA expression. In the last decades, the miniaturization has worsened this situation, showing dramatic divergences between KM-RTA predictions and the experimental results when bulk and nanoscale samples are simulated with the same relaxation time expressions.

In contrast, when the system is not so close to equilibrium, VM provides a better way to solve BTE. In general the collision terms in VM cannot be expressed analytically, instead they have to be obtained by integration using a trial function. This trial function should be close to the actual solution to have a good convergence.
The main drawback is that this function is not necessarily the same in all temperature ranges. In conclusion, this approach is only useful in regions where the form of the phonon distribution is known to some extent.

Although thermal conductivity obtained within the KM and VM seem to be disconnected, from thermodynamic reasoning we will demonstrate that both can be derived from the balance of entropy production. The main difference between both approaches resides in the way this balance is performed. Starting from this point, it is easy to demonstrate that a general expression for thermal conductivity can be obtained by combining the distribution function in these two extreme situations: the first one where resistive processes are dominant and equilibrium can be rapidly achieved (related to KM) and the second one where although equilibrium cannot be easily reached, conservation of momentum in collisions allows us to determine analytically the scattering term (related to VM).

To obtain $\kappa$ in each limit, an expression for the scattering term in Eq. (\ref{boltz_0}) is needed. In KM this is usually done by the RTA approach, but in VM the expected form of the distribution function do not provide a simple expression. In the particular case when normal collisions are dominant, we suggest that the same RTA expression can be used, leaving the difference between approaches only in the way to perform the thermodynamic averages with this relaxation times.

\subsection{Resistive vs Normal scattering (equilibrium vs non-equilibrium)}
\label{resistive_normal}

As indicated before, the expected form of $\Phi_{\bm{q}}$ will determine the choice between a KM or a VM approach. The calculation of the scattering rates depends on it and, at the same time this depends on which scattering mechanism is dominating the system. Determining the dominating mechanism is thus the first important question to solve.

Phonons can relax by different mechanisms, colliding with boundaries, impurities, electrons and between them. All these mechanisms are resistive except some part of the phonon-phonon collisions. Two phonons with wave number and energy $(\bm q,\omega_{\bm q})$ and $(\bm q',\omega_{\bm q'})$ can scatter and produce, as a result, a new phonon $(\bm q'',\omega_{\bm q''})$ (or vice-versa). In all events, energy must be conserved, but the wave number or quasi-momentum can be lost due to the interaction with the whole lattice. The equation
\begin{equation}
\bm q+\bm q'=\bm q''+\bm G,
\end{equation}
where $\bm G$ is a reciprocal lattice vector, expresses the fact that the total lattice can acquire an amount of momentum $\bm G$ because the resultant phonon is reflected outside of the first Brillouin zone (BZ).\cite{Ziman1979} If the quasi-momentum is conserved ($\bm G=0$) the scattering processes are called normal or N-processes, while in the general case ($\bm G\neq 0$) they are called Umklapp or U-processes. Regarding the dominance of the N-processes two limiting behaviors can be considered:

i) When resistive collisions are dominant and N-processes are negligible, momentum will be completely dissipated and its average value is zero. The only way to move the phonon distribution from equilibrium is by changing its temperature. In that case, the distribution function takes the form
\begin{equation}
\label{dist_rta}
f_{\bm{q}}=\frac{1}{e^{\hbar \omega_{\bm{q}}/k_{B}(T+\delta T)}-1}\approx
\frac{1}{e^{\hbar \omega_{\bm{q}}/k_{B}T}e^{1-\delta T/T}-1}.
\end{equation}
Comparing with Eq. (\ref{aprox_non_eq_dist}) an expression for $ \Phi_{\bm{q}}$ can be obtained
\begin{equation}\label{phi_for_rta}
 \Phi_{\bm{q}}= \hbar \omega_{\bm{q}} \frac{\delta T}{T}.
\end{equation}
In this situation KM is the most suitable approach to use.

ii) When N-processes are dominant, the system will not be able to relax the momentum to zero (the quasi-momentum is conserved) and a displacement $\bm u$ of the distribution function in the direction of the thermal gradient is expected. The distribution function takes the form \cite{Krumhansl1965}
\begin{equation}
\label{dist_vm}
f_{\bm{q}}=\frac{1}{e^{(\hbar \omega_{\bm{q}}-\bm{u}\cdot {\bm{q}})/k_{B}T}-1}
\end{equation}
which is in a non-equilibrium situation. Then, $\Phi_{\bm{q}}$ takes the form
\begin{equation}\label{phi_for_vm}
 \Phi_{\bm{q}}=\bm{u}\cdot \bm{q}
\end{equation}
In this case the VM approach must be used.

Summing up, Eqs. (\ref{phi_for_rta}) and (\ref{phi_for_vm}) are the two forms of $ \Phi_{\bm{q}}$ expected for the distribution function in each approach, KM and VM respectively, corresponding to two extreme situations described above. Next, we will use both expressions of $ \Phi_{\bm{q}}$ to show that in some situations, they yield equivalent expressions for the relaxation times.

\subsection{Defining Scattering rates}

Once we have determined both expressions for $\Phi_{\bm{q}}$ in the two limiting cases, we can use them to determine the collision term in Eq. (\ref{boltz_0}) in each case. This depends on the transition probabilities and the form of the distribution functions. In general, the collision term in the Boltzmann equation can be written, for elastic scattering, as
\begin{equation}
\label{collision_1}
\left.\frac{\partial f_{\bm{q}}}{\partial t}\right|_{\textrm{\tiny{scat}}}=\int \frac{\Phi_{\bm{q}}-\Phi_{\bm{q}'}}{k_{B}T} P_{\bm{q}}^{\bm{q}'} d\bm{q}'.
\end{equation}
where $P_{\bm{q}}^{\bm{q'}}$ are the scattering transition rates from mode $\bm{q}$ to $\bm{q'}$ when the distribution functions correspond to equilibrium.

The integral (\ref{collision_1}) is expressing the fact that relaxation process in an out of equilibrium is modified by terms $\Phi_{\bm{q}}-\Phi_{\bm{q'}}$, \textit{i. e.} depending on the displacement with respect to equilibrium of the different colliding particles. Expression (\ref{collision_1}) can be generalized for an arbitrary number of colliding particles:
\begin{widetext}
\begin{equation}
\label{collision_2}
\left.\frac{\partial f_{\bm{q}}}{\partial t}\right|_{\textrm{\tiny{scat}}} = \frac{1}{k_BT}
\int \left[\Phi_{\bm{q}} + \displaystyle\sum_{i=1}^{n}\Phi_{\bm{q}_{i}}-\displaystyle\sum_{j=1}^{m}
\Phi_{\bm{q}_{j}}\right] P_{\bm{q}\bm{q}_{1}...\bm{q}_{n}}^{\bm{q}'_{1}..\bm{q}'_{m}}
\prod_{i=1\atop j=1}^{m\atop n} d\bm{q}_{i} d\bm{q}'_{j}
\end{equation}
\end{widetext}
where $\bm{q}$ collides with $\left\{\bm{q}_{i}\right\}$ giving as a result the modes $\left\{\bm{q}'_{j}\right\}$.
Expression (\ref{collision_2}) shows the main complexity of solving the BTE equation. The scattering term requires the actual distribution function inside an integral expression establishing BTE as an integro-differential equation. One can use a numerical approach to solve it but other approximations can also be employed. These are usually based in the fact that the distribution used in the integral does not modify significantly the final result in some limiting situations.
In RTA we assume that the system is close enough to equilibrium that the differences between using the actual form of the distribution or the equilibrium form in the collision integral (\ref{collision_2}) is not significant.This is like saying that the only mode out of equilibrium is the one with wave number $\bm{q}$ and that the remaining modes rest in equilibrium. Thus
\begin{equation}\label{condition_1}
\Phi_{\bm{q}_{i}}=\Phi_{\bm{q}'_{j}}=0
\end{equation}
for all $\bm{q}_{i}\neq\bm{q}$ and $\bm{q}'_{j}\neq\bm{q}$. In this case,
\begin{equation}\label{collision_3}
\left.\frac{\partial f_{\bm{q}}}{\partial t}\right|_{\textrm{\tiny{scat}}} = \frac{\Phi_{\bm{q}}}{k_{B}T}
\int P_{\bm{q}\bm{q}_{1}...\bm{q}_{n}}^{\bm{q}'_{1}..\bm{q}'_{m}}\prod_{i=1\atop j=1}^{m\atop n} d\bm{q}_{i} d\bm{q}'_{j}
\end{equation}
If we substitute Eq. (\ref{aprox_non_eq_dist}) in (\ref{collision_3}) we have
\begin{equation}\label{collision_4}
\left.\frac{\partial f_{\bm{q}}}{\partial t}\right|_{\textrm{\tiny{scat}}} = \frac{f_{\bm{q}}-f^0_{\bm{q}}}{f_{\bm{q}}^{0}(f_{\bm{q}}^{0}+1)}
\int P_{\bm{q}\bm{q}_{1}...\bm{q}_{n}}^{\bm{q}'_{1}..\bm{q}'_{m}}\prod_{i=1\atop j=1}^{m\atop n} d\bm{q}_{i} d\bm{q}'_{j} \quad .
\end{equation}

Thus, we can define the relaxation time $\tau_{\bm{q}}$ of mode $\bm{q}$ as
\begin{equation}\label{collision_5}
\frac{1}{\tau_{\bm{q}}} = \frac{1}{f_{\bm{q}}^{0}(f_{\bm{q}}^{0}+1)} \int P_{\bm{q}\bm{q}_{1}...\bm{q}_{n}}^{\bm{q}'_{1}..\bm{q}'_{m}} \prod_{i=1\atop j=1}^{m\atop n} d\bm{q}_{i} d\bm{q}'_{j}
\end{equation}
and so, we obtain the BTE solution in the well known RTA approach
\begin{equation}\label{RTA}
\left.\frac{\partial f_{\bm{q}}}{\partial t}\right|_{\textrm{\tiny{scat}}} = \frac{f_{\bm{q}}-f_{\bm{q}}^{0}}{\tau_{\bm{q}}} \quad .
\end{equation}

We can make a similar assumption when normal scattering is the dominant relaxation process. The only change is that the distribution function where the actual distribution function will relax is that of Eqs. (\ref{dist_vm})-(\ref{phi_for_vm}). In that case, condition (\ref{condition_1}) cannot be fulfilled locally by each mode, but it can be demonstrated that in the linear regime not much error is made in Eq. (\ref{collision_2})\cite{Ziman1979} if we consider that 
\begin{equation}\label{collision_6}
\int \Phi_{\bm{q}_{i}} P_{\bm{q}\bm{q}_{1}...\bm{q}_{n}}^{\bm{q}'_{1}..\bm{q}'_{m}} d\bm{q}_{i}=
\int \Phi_{\bm{q}'_{j}} P_{\bm{q}\bm{q}_{1}...\bm{q}_{n}}^{\bm{q}'_{1}..\bm{q}'_{m}} d\bm{q}_{i}=0 \qquad\forall i,j
\end{equation}
This condition leads to the same result as that obtained near equilibrium (\ref{collision_4})-(\ref{collision_5}), since condition (\ref{collision_6}) is equivalent to condition (\ref{condition_1}). Thus, we can use the same expression for the scattering rates in both limiting situations, near equilibrium and in non-equilibrium, despite of the very different nature of the two situations.
Note that by using Eqs.  (\ref{dist_rta})-(\ref{phi_for_rta}) and (\ref{dist_vm})-(\ref{phi_for_vm}) in eq. (\ref{collision_6}) we are not stating that resistive processes are suppressed. In fact $P_{\bm{q}\bm{q}_{1}...\bm{q}_{n}}^{\bm{q}'_{1}..\bm{q}'_{m}}$ are the transition rates for all the resistive scattering processes. We are only considering that collision integral (\ref{collision_6}) does not change significantly when one uses the actual form of the distribution function or the proposed approximations in the corresponding regimes.

In RTA in the special case when all the resistive terms are absolutely negligible, either KM and VM expression will give an infinite thermal conductivity as expected. In the following section we apply this result to obtain an expression for the thermal conductivity under each situation in the case where non-negligible resistive terms are present. The approximations here proposed will allow us to calculate two well differentiate regimes of behavior in the thermal transport: the kinetic and the collective regime.

\section{Thermal conductivity regimes}
Here we propose to derive thermal conductivity from the balance of entropy as obtained by Ziman\cite{Ziman1979}. The reason for this choice lies in the nature of normal scattering. Entropy generation is related to resistive collisions and normal scattering is not resistive. It is logical to think that entropy production can be modified when these kind of collisions are dominant. In this section we analyze these differences.

In this deduction thermal conductivity is obtained from the equality of entropy production calculated from the drift and the collision terms in Eq. (\ref{boltz_0}). The collision term is obtained under a microscopic formalism, and the drift term is expressed in thermodynamic variables. 

The key point to notice is that N-processes, despite of being non-resistive, mix the different modes, affecting the balance between drift and collisions. If N-processes are not important and mode mixing is low, entropy balance should be fulfilled individually by each mode, that is, locally in momentum space. This leads to the thermal conductivity in the \textit{kinetic regime}. On the other hand, when mode mixing is high (N-processes dominate) the entropy balance should be achieved globally, in this case we obtain the thermal conductivity in the \textit{collective regime}. Depending on the intensity of the normal collisions we should select the local or the global version for the entropy production balance. Next, we detail both regimes of behavior and obtain the corresponding thermal conductivity contribution.

\begin{figure}[htb]
\subfloat[]{\includegraphics[width=0.4\textwidth]{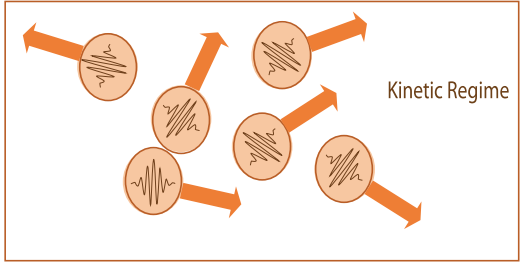}}\\
\subfloat[]{\includegraphics[width=0.4\textwidth]{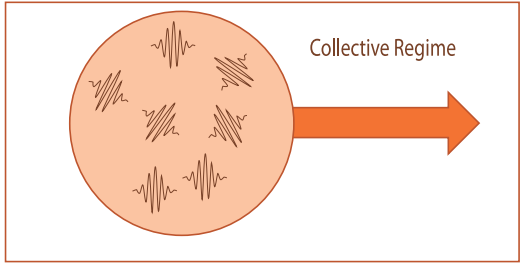}}
\caption{(Color online) These schemes illustrate the behavior of the phonons in each regime: (a) In the kinetic regime N-processes are negligible. The phonon distribution is near equilibrium and resistive scatterings tend to bring it back to equilibrium. Each phonon contributes independently to the heat flux and so the equation of the entropy balance must be fulfilled individually by each mode. (b) In the collective regime N-processes dominate and the distribution is in non-equilibrium. Momentum is conserved and shared among the phononic modes through N-processes. The phonons behave as a collectivity rising a total heat flux and so the equation of the entropy balance must be fulfilled globally.\label{fig:regimes}}
\end{figure}

\subsection{Kinetic regime}
Entropy of a distribution of bosons is
\begin{equation}
\label{entropy}
\frac{s_{\bm{q}}}{k_{B}}= f_{\bm{q}} \ln f_{\bm{q}} - (f_{\bm{q}}-1) \ln (f_{\bm{q}}-1) \quad .
\end{equation}
The variation of entropy can be obtained from Eqs. (\ref{entropy}) and (\ref{aprox_non_eq_dist}). If we take only linear terms in $\Phi_{\bm{q}}$ this can be written as \cite{Ziman1979}
\begin{equation}
\label{entropy_production_local}
\left.\dot{s}_{\bm{q}}\right|_{\textrm{\tiny{scat}}}=\left.\frac{\partial s_{\bm{q}}}{\partial t}\right|_{\textrm{\tiny{scat}}} = \frac{\Phi_{\bm{q}}}{T}\left.\frac{\partial f_{\bm{q}}}{\partial t}\right|_{\textrm{\tiny{scat}}},
\end{equation}
Thermodynamically, the entropy variation can be also written in terms of the heat flux
\begin{equation}
\label{entropy_drift_def}
\left.\dot{s}_{\bm{q}}\right|_{\textrm{\tiny{drift}}}=\left.\frac{\partial s_{\bm{q}}}{\partial t}\right|_{\textrm{\tiny{drift}}} = {\bm j}_{\bm{q}}\cdot \nabla \left(\frac{1}{T}\right) =
\frac{{\bm j}^{2}_{\bm{q}}}{\kappa_{\bm{q}} T^2}
\end{equation}
where the heat flux of mode $\bm q$ is
\begin{equation}
\label{flux_micro}
{\bm j}_{\bm{q}}= \hbar \omega_{\bm q}\bm{v}_{g} (f_{\bm{q}}-f^{0}_{\bm{q}}) = \hbar \omega_{\bm q} \bm{v}_{g} f_{\bm{q}}^{0}(f_{\bm{q}}^{0}+1)\frac{ \Phi_{\bm{q}}}{k_BT}
\end{equation}
and we have used the fact that $\bm{j}_{\bm{q}}=-\kappa_{\bm{q}} \nabla T$, where $\kappa_{\bm{q}}$ is the thermal conductivity of mode $\bm{q}$, and $\bm{v}_{g}$ is the group velocity.

Equating (\ref{entropy_drift_def}) and (\ref{entropy_production_local}) leads to an expression giving the thermal conductivity of each mode
\begin{equation}
\label{conductivity_mode_kin}
\kappa_{\bm{q}}=\frac{{\bm j}^{2}_{\bm{q}}}{T\Phi_{\bm{q}}\left.\frac{\partial f_{\bm{q}}}{\partial t}\right|_{\textrm{\tiny{scat}}}}.
\end{equation}
Integrating (\ref{conductivity_mode_kin}) over all modes yields total thermal conductivity in this kinetic regime
\begin{equation}
\kappa_{\rm{kin}}=\int\kappa_{\bm{q}}d{\bm{q}}=\int\frac{{\bm j}^{2}_{\bm{q}}}{T\Phi_{\bm{q}}\left.\frac{\partial f_{\bm{q}}}{\partial t}\right|_{\textrm{\tiny{scat}}}}d{\bm{q}}
\end{equation}
and if we substitute Eq. (\ref{flux_micro}) we finally obtain
\begin{equation}
\label{conductivity_kin}
\kappa_{\rm{kin}}=\int\frac{\left[\hbar \omega_{\bm q} \bm{v}_{g} f_{\bm{q}}^{0}(f_{\bm{q}}^{0}+1)\frac{ \Phi_{\bm{q}}}{k_BT}\right]^2}{T\Phi_{\bm{q}}\left.\frac{\partial f_{\bm{q}}}{\partial t}\right|_{\textrm{\tiny{scat}}}}d{\bm{q}}
\end{equation}

\subsection{Collective regime}
In the second limiting case, phonons behave as a collectivity and each mode do not contribute to the entropy production individually but collectively. In this case the balance of entropy should be achieved globally and integration should be performed before equating terms.
 Thus, the total entropy production is on one side
 \begin{equation}
 \label{entropy_global_scat}
 \left.\dot{s}_{\rm{tot}}\right|_{\textrm{\tiny{scat}}} =\int \left. \dot{s}_{\bm{q}}\right|_{\textrm{\tiny{scat}}}d{\bm{q}} = \int \frac{\Phi_{\bm{q}}}{T}\left.\frac{\partial f_{\bm{q}}}{\partial t}\right|_{\textrm{\tiny{scat}}} d{\bm{q}}
 \end{equation}
 and on the other we must account for a total heat flux, giving
\begin{equation}
\left.\dot{s}_{\rm{tot}}\right|_{\textrm{\tiny{drift}}} = {\bm j^2_{\rm{tot}}} \cdot \nabla \left(\frac{1}{T}\right).
\end{equation}
Using the Fourier's law ${\bm j_{\rm{tot}}} = -\kappa\nabla T$, we obtain
\begin{equation}
\label{entropy_global_drift}
\left. \dot{s}_{\rm{tot}} \right|_{\textrm{\tiny{drift}}} = \frac{\bm j^2_{\rm{tot}}}{\kappa T^2}.
\end{equation}
being $\kappa$ the global thermal conductivity achieved in this regime. We denote it as $\kappa_{\rm{coll}}$ and we obtain its expression by equating (\ref{entropy_global_scat}) and (\ref{entropy_global_drift})
\begin{equation}
\label{conductivity_global}
\kappa_{\rm{coll}}=\frac{\bm j^2_{\rm{tot}}}{T^2\int \frac{\Phi_{\bm{q}}}{T}\left.\frac{\partial f_{\bm{q}}}{\partial t}\right|_{\textrm{\tiny{scat}}}d{\bm{q}}}
\end{equation}
where the total heat flux is
\begin{equation}\label{heat_flux_collect}
{\bm j_{\rm{tot}}}=\int {\bm j}_{\bm{q}}d{\bm{q}}=\int  \hbar \omega_{\bm q} \bm{v}_{g} f_{\bm{q}}^{0}(f_{\bm{q}}^{0}+1)\frac{ \Phi_{\bm{q}}}{T}d{\bm{q}}
\end{equation}

By substituting this expression in Eq. (\ref{conductivity_global}), we have

\begin{equation}
\label{conductivity_global2}
\kappa_{\rm{coll}}=\frac{\left[\int  \hbar \omega_{\bm q} \bm{v}_{g} f_{\bm{q}}^{0}(f_{\bm{q}}^{0}+1)\frac{ \Phi_{\bm{q}}}{k_BT}d{\bm{q}} \right]^2}{T^2\int \frac{\Phi_{\bm{q}}}{T}\left.\frac{\partial f_{\bm{q}}}{\partial t}\right|_{\textrm{\tiny{scat}}} d{\bm{q}} }
\end{equation}
This new regime relies on a thermodynamic basis, and it can not be deduced from a framework where normal scattering is treated as a resistive mechanism like in Callaway model.
After deducing the expression of the thermal conductivity in each regime, we need to choose a magnitude able to determine if we are in the local or global behavior. Secondly in order to calculate the integrals in (\ref{conductivity_kin}) and (\ref{conductivity_global2}), we need some expressions for the collision terms. This will be done in the next section.

\section{Thermal conductivity in terms of frequency and relaxation times}\label{sec:thermal_conduct}

We are now able to calculate the thermal conductivity from Eq. (\ref{conductivity_kin}) for the kinetic regime and from Eq. (\ref{conductivity_global2}) for the collective regime. In order to obtain numerical results, first we need to express them in terms of the equilibrium distribution function and the relaxation times. Using (\ref{aprox_non_eq_dist})-(\ref{RTA}) in Eq. (\ref{conductivity_kin}), $\kappa_{\rm{kin}}$ can be rewritten as
\begin{equation}
\label{conductivity_kin_2}
\kappa_{\rm{kin}}=\int \hbar \omega_{\bm{q}} \tau_{\bm{q}} \bm{v}_g^{2} \frac{\partial f_{\bm{q}}^{0}}{\partial T} d{\bm{q}}
\end{equation}
which is the classical KM expression for the thermal conductivity. Here and onward we have omitted the index for the phonon branch in the integrals for the shake of simplicity. 

For (\ref{conductivity_global2}), one can make the same substitutions to obtain
\begin{equation}
\label{conductivity_global_3}
\kappa_{\rm{coll}}=\frac{\left(\int \Phi_{\bm{q}} \bm{v}_{g} \frac{\partial f_{\bm{q}}^{0}}{\partial T} d{\bm{q}}\right)^{2}}{\int \frac{\Phi_{\bm{q}}^{2}}{\hbar \omega_{\bm{q}}}\frac{1}{\tau_{\bm{q}}} \frac{\partial f_{\bm{q}}^{0}}{\partial T}   d{\bm{q}} } \quad.
\end{equation}

$\kappa_{\rm{kin}}$ and $\kappa_{\rm{coll}}$ can be re-expressed in terms of frequency to simplify the integration in isotropic materials. This is done by the substitution  $d{\bm{q}}\rightarrow D_{\omega}d\omega$ , being $D_{\omega}$ the density of states (DOS), and integrating the angular part. For the kinetic regime this leads to the expression
\begin{equation}
\label{conductivity_kin_dos}
\kappa_{\rm{kin}}=\frac{1}{3}\int \hbar \omega \tau_{\omega} v_g^{2} \frac{\partial f^0_{\omega}}{\partial T} D_{\omega}d\omega
\end{equation}
where now the frequency dependence  is indicated with the subindex (in the group velocity the subindex is omitted for simplicity), and  for the collective regime
\begin{equation}\label{conductivity_coll_dos}
\kappa_{\mathrm{coll}}=\frac{1}{3}\frac{\left(\int v_{g} q_{\omega} \frac{\partial f^0_{\omega}}{\partial T}D_{\omega}d\omega\right)^{2}}{\int \frac{q^2_{\omega}}{\hbar \omega}\frac{1}{\tau_{\omega}} \frac{\partial f^0_{\omega}}{\partial T}D_{\omega}d\omega}
\end{equation}
where we have used the explicit form (\ref{phi_for_vm}) to express $\Phi_{\bm{q}}$ in terms of the wave vector $q_{\omega}$. The only question to be addressed in Eq. (\ref{conductivity_coll_dos}) is that in order to maintain isotropy, $q^2_{\omega}$ should be a frequency averaged value. This does not lead to large variations in isotropic materials.

As we have already pointed, in both expression (\ref{conductivity_kin_dos}) and (\ref{conductivity_coll_dos}) $\tau_{\omega}$ is the same and accounts for the total relaxation time contributing to thermal resistance. Then, we denominate it  $\tau_{R_\omega}$.
Finally, we need a magnitude which accounts for the kind of regime the phonon distribution is undergoing at the different temperatures. As we have commented, this is determined by the degree of mixing between modes. Since this is related to the dominance of normal with respect to resistive processes, a switching factor weighting the relative importance of these processes should be used. This factor can be calculated from a matrix representation\cite{Guyer1966a}
\begin{equation}\label{sigma}
\Sigma\equiv\frac{1}{1+\frac{<\tau_N>}{<\tau_R>}}
\end{equation}
where $\tau_N$ is the relaxation time due to N-processes and $\tau_R$ is the relaxation time due to resistive processes. Both relaxation times $\tau_N$ and $\tau_R$ are averaged over all modes. This is calculated as
\begin{equation}\label{k_kin_int}
\langle \tau_{i}\rangle=\frac{\int \hbar \omega \tau_{i_\omega} \frac{\partial f^0_{\omega}}{\partial T}d\bm{q}}{\int \hbar \omega \frac{\partial f^0_{\omega}}{\partial T}d\bm{q}}
\end{equation}
with subindex $i$ indicating $N$ or $R$.

The general expression of the thermal conductivity must include this switching factor to account for all the intermediate regimes between the limiting regimes, \textit{i}. \textit{e}. from kinetic to collective regime. Thus,
\begin{equation}
\label{kappa_tot}
  \kappa=\kappa_{\mathrm{kin}}(1-\Sigma)+ \kappa_{\mathrm{coll}}\Sigma
\end{equation}

If we are in the kinetic (unmixed-mode) limit $\tau_N>>\tau_R$ then $\Sigma \rightarrow 0$ and  $\kappa\rightarrow\kappa_{\mathrm{kin}}$. If we are in the collective (mixed-mode) limit  $\tau_N<<\tau_R$ then $\Sigma \rightarrow 1$ and $\kappa\rightarrow\kappa_{\mathrm{coll}}$.

Different phenomenological behavior can be deduced from the mathematical difference in performing the averages in (\ref{conductivity_kin_dos}) and (\ref{conductivity_coll_dos}). This differences are equivalent to add resistivities in serial or parallel, if we interpret the scattering events on a particular mode as a resistance. This can give physical insight in order to interpret the thermal conductivity behavior in the different regimes.
From Eq. (\ref{kappa_tot}) it can be deduced why all models based on a single approach (KM or VM) fail when extended to a global model in a large range of temperatures. In this extension they are used in an approximation where they are not supposed to be valid. With this Eq. (\ref{kappa_tot}), the behavior change is included in the model, extending its applicability to the whole temperature range. Another remarkable difference is the way to include size effects in both expressions. This is discussed on the next section.

\subsection{Size-effects on the kinetic and collective terms}

In an infinite semiconductor sample at near room temperature one can consider that only impurities scattering and umklapp scattering participate significantly, then by means of the Mathiessen's rule
\begin{equation}\label{mat_bulk}
\tau^{-1}_{R_\omega}=\tau^{-1}_{I_\omega}+\tau^{-1}_{U_\omega} \quad.
\end{equation}

Relaxation times allow to calculate a related term, the phonon mean free path $\ell$, that is the product between the relaxation time of a mode and its group velocity $\ell=v_g\tau$. If the dimension of the system is finite and the temperature is low, intrinsic mean free paths can be larger than the size of the system. In this case, boundary effects need to be included.

In the kinetic regime of the thermal conductivity, as the phonons behave individually, each mode could experience independently a scattering with the boundary. Then, an extra term considering this effect should be included in the kinetic term of Eq. (\ref{kappa_tot}) by using the Mathiessen's rule in combination with the intrinsic events, this is $\tau_{B_\omega}$ the relaxation time due to boundary scattering
\begin{equation}\label{mat_boundary}
\tau^{-1}_{R_\omega}=\tau^{-1}_{I_\omega}+\tau^{-1}_{U_\omega}+\tau^{-1}_{B_\omega} \quad .
\end{equation}

However, in the collective term some caution has to be taken. In this regime a scattering rate is a quantity describing the distribution globally. In other words, one cannot assume an extra scattering term in each mode independently because the boundary is noticed by the whole phonon collectivity. Thermodynamically, this is the same situation as flow on a pipe. Carriers in the center of the pipe notice the boundary not by themselves but through the collisions with the rest of the particles. The net effect on the flow is the reduction of the flow on the surface. The usual solution for this situation is to assume that the flow on the surface is zero. This is feasible if surfaces are rough enough. Once imposed this extra assumption, a geometrical factor $F$ depending on the roughness and the transversal size of the system should be included in the collective term of Eq. (\ref{kappa_tot}). In the work by Guyer and Krumhansl\cite{Guyer1966a} this factor is calculated for a cylindrical shape. In order to generalize the geometrical factor to account for several geometries and so extend the range of validity of the collective term from bulk to small size samples, we used an expression derived in a previous work\cite{Xavi2007}
\begin{equation}
F(L_{\rm eff})=\frac{1}{2\pi^2}\frac{L_{\rm eff}^2}{\ell^2}\left(\sqrt{1+4\pi^2\frac{\ell^2}{L_{\rm eff}^2}}-1\right)\label{factor_F}\quad ,
\end{equation}
being $\ell$ the phonon mean free path and $L_{\rm eff}$ is the effective length of the system. By geometrical considerations it can be deduced\cite{Ziman1979,Zhang2007} that $L_{\mathrm{eff}}=d$ for nanowires of diameter $d$, $L_{\mathrm{eff}}=\sqrt{\pi/2}L$ for square wires of size $L$ and $L_{\mathrm{eff}}=2.25h$ for thin layers of thickness $h$. Expression (\ref{factor_F}) was obtained in the framework of the Extended Irreversible Thermodynamics\cite{EIT1993} and includes in its derivation higher order terms into the BTE expansion, which can be important when the size of the samples are of the order of the phonon mean free path and it has some advantages: it is analytical, it can be used for different geometries and it takes automatically into consideration the degree of non-equilibrium present in the sample depending on the normal and resistive relaxation times. Regarding the mean free path, from the works by Alvarez \textit{et al.}\cite{Xavi2007} and Guyer-Krumhansl\cite{Guyer1966a} it can be easily deduced that $\ell=v_g\sqrt{\langle\tau_N\rangle\langle\tau_R^{-1}\rangle^{-1}}$, reminding that mean relaxation times are calculated from Eq. (\ref{k_kin_int}).

Finally, the thermal conductivity for small size samples would be
\begin{equation}\label{kappa_final}
\kappa=\kappa_{\rm{kin}}(1-\Sigma)+\kappa_{\rm{coll}}\Sigma F(L_{\rm eff})
\end{equation}
Note that if $\ell/L_{\mathrm{eff}}\rightarrow 0$ ($\ell \ll L_{\mathrm{eff}}$), then $F(L_{\rm eff})\rightarrow 1$ and we recover Eq. (\ref{kappa_tot}). In the opposite limit, $\ell/L_{\mathrm{eff}}\rightarrow \infty$ ($\ell \gg L_{\mathrm{eff}}$), $F\sim L_{\mathrm{eff}}/\pi\ell\rightarrow 0$.

Next, we will test the validity of our model by applying it on different silicon samples since Si is a well-characterized semiconductor in the literature. This requires to calculate previously its dispersion relations, DOS, and relaxation times.

\section{Silicon dispersion relations and density of states}

The bond charge model (BCM) proposed by Weber \cite{Weber1974} provides accurate and complete dispersion relation for group IV semiconductors, III-V and II-VI compounds, and they can be obtained with a minimum set of force constants, actually 4 parameters for Si, and 5 in the case of III-V or II-VI compounds.\cite{Camacho1999} Furthermore, the BCM reproduces very well the transversal acoustic phonon branches close to the border of the Brillouin zone while other models with much more parameters are not able to do it. The use of the complete dispersion relation includes the role of optical phonons on $\kappa$, neglected in the Debye approximation.
\begin{figure}[htb]
\subfloat[]{\includegraphics[width=0.43\textwidth]{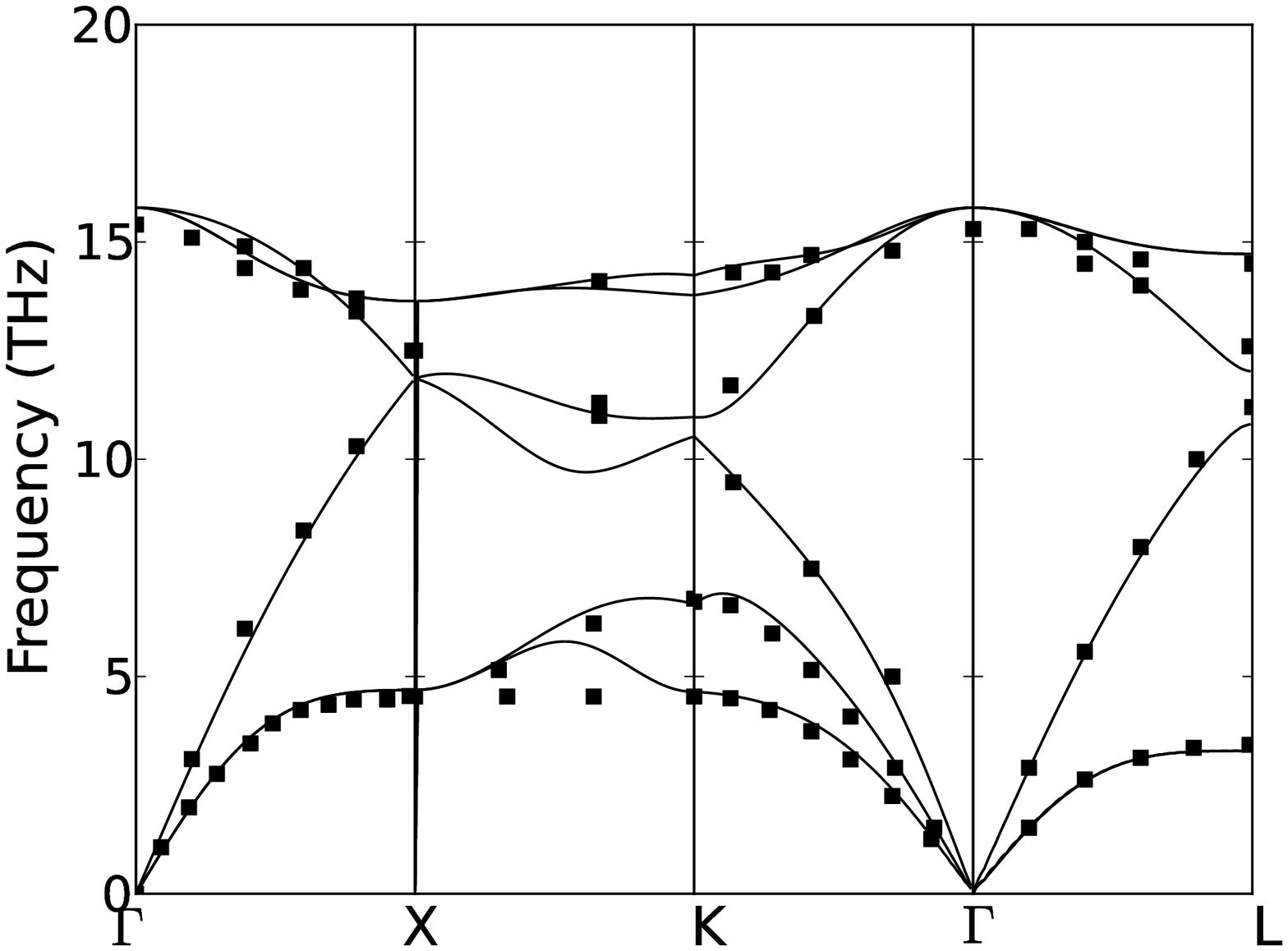}}\\
\subfloat[]{\includegraphics[width=0.4\textwidth]{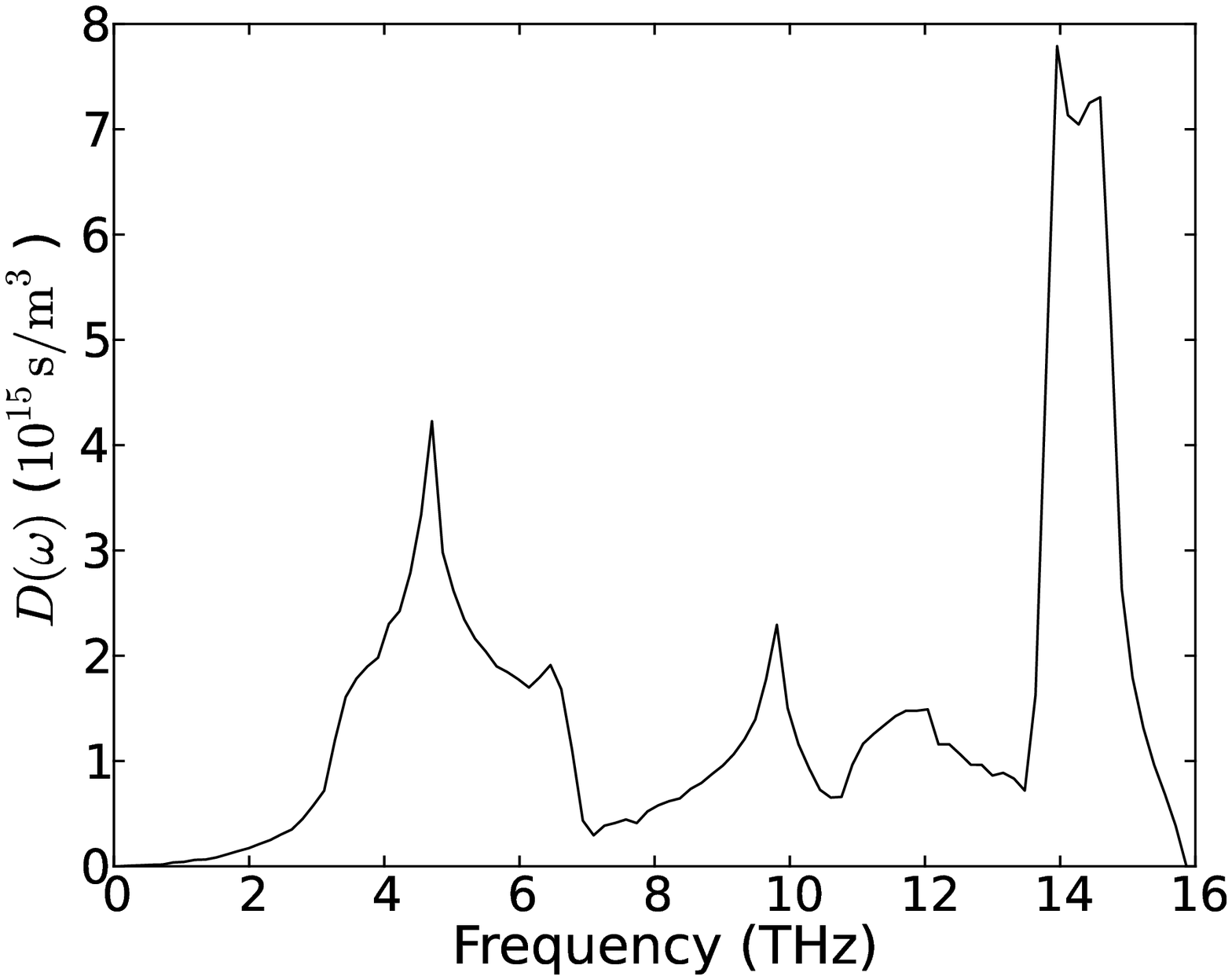}}
\caption{(a) Phonon dispersion relations for Si along some high-symmetry directions calculated with the BCM (lines) compared to neutron experimental data\cite{Nilsson1972} (symbols). (b) DOS calculated for Si using the BCM.\label{fig:RD}}
\end{figure}
In Fig. \ref{fig:RD} we show the dispersion relations and DOS we have computed with the BCM. The dispersion relations fit very well the neutron experimental data,\cite{Nilsson1972} as the \textit{ab initio} calculations performed by Ward and Broido.\cite{Broido2010} The DOS calculation agrees also very well with the literature.

\section{Silicon relaxation times: dependence with frequency and temperature}

Expressions for the relaxation times are also needed to compare with data. We have chosen simple expressions in order to show that even in this case they lead to remarkable predictions. The use of more accurate expressions obtained for example by \textit{ab initio} calculations will lead for sure to better fits.

In the case of impurity scattering (or mass defect), we use the expression
\begin{equation}
\label{tauI}
\tau^{-1}_{I_\omega}=\frac{\pi}{6}V\Gamma\omega^2 D_{\omega}\quad ,
\end{equation}
being $V$ the atomic volume and $\Gamma= \sum_i f_i\left(\Delta M/M\right)^2 $ the mass-fluctuation factor, with $f_i$ the isotopic fraction. This expression is given by Tamura\cite{Tamura1983} and it is obtained from second-order perturbation theory for diamond-like materials. This general expression recovers the conventional expression given by Klemens\cite{Klemens1955} under Debye model conditions , \textit{i.e.} $\tau^{-1}_{I_\omega}=A\omega^{4}$ with $A=V\Gamma/(4\pi v_{g}^{3})$.  The advantage of Eq.  (\ref{tauI}) is that gives us a calculated relaxation time, with no fitting parameters.

For boundary scattering, we use the usual expression\cite{Casimir1938,Berman1953,Berman1955}
\begin{equation}
\tau^{-1}_{B_\omega}=\frac{v_{g}}{L_{\mathrm{eff}}}\quad ,
\label{tauB}
\end{equation}
where $L_{\mathrm{eff}}$ is the effective length of the sample and $v_{g}$ again the group velocity calculated from the dispersion relations.

The relaxation times for N- and U-processes will be taken, in the intermediate temperature range, from those provided by Ward and Broido,\cite{Broido2010} which fit their \textit{ab initio} calculations. We have modified their expressions of $\tau_{U}$ and $\tau_{N}$ in order to extend them to the low and high temperature regimes, respectively. As shown by Herring,\cite{Herring1954} N-scattering at low temperatures must be of the form $\omega^{n} T^{5-n}$, $n$ being an integer, while at the high temperature region it should follow a $T^{-1}$ law. Since the expression provided by Ward and Broido does not follow the right temperature dependence at high temperatures, we have included the additional term  $1/(B' _{N}T)$. In this way, the expression will be valid in the whole temperature range
\begin{equation}
\tau_{N_\omega}=\frac{1}{B' _{N}T}+\frac{1}{B_{N}T^{3}\omega^2[1-\exp(-3T/\Theta_D)]}\quad .
\label{tauN}
\end{equation}
where $\Theta_D$ is the Debye temperature. Concerning U-processes, following the argument provided by Ziman,\cite{Ziman1979} at low temperatures the scattering of two phonons with wave vectors $\bm q_1$ and $\bm q_2$ cannot provide $\bm q_3+\bm G$, with $\bm G\neq 0$, since low temperature means low energy or low $\bm q_i$. In other words, U-processes are not possible at low temperature. We have established a temperature limit assuming that, for $\bm q_U=2\pi/3a$ ($a$ being the lattice parameter of Si), $1/3$ the limit of the Brillouin zone, the probability of U-processes decreases exponentially. The temperature limit $\Theta_U$ is calculated through the expression $\hbar\omega_{\bm q_U}\approx k_B\Theta_U$. For Si we obtain $\Theta_U\approx 100$ K.  The final expression for U-processes is:
\begin{equation}
\tau_{U_\omega}=\frac{\exp(\Theta_U/T)}{B_U \omega^4T[1-\exp(-3T/\Theta_D)]}\quad .
\label{tauU}
\end{equation}
At high enough temperatures, the numerator of Eq. (\ref{tauU}) is 1 and we recover Ward and Broido's expression.

\section{Results and Discussion}
\begin{figure}[htb]
\includegraphics[width=0.45\textwidth]{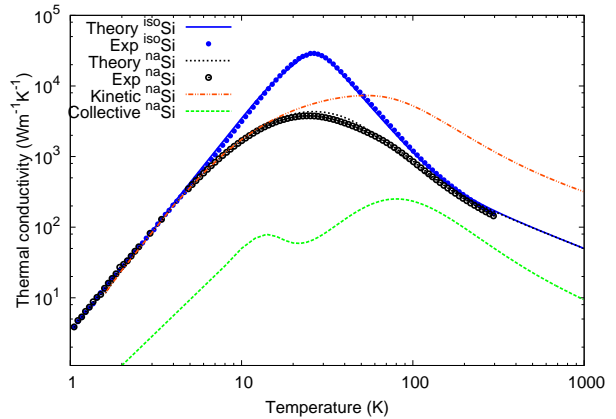}
\caption{(Color online) Total thermal conductivity as a function of temperature in a double logarithmic plot for $^{\rm na}$Si and $^{\rm iso}$Si, as a result of fitting Eq. (\ref{kappa_final}) to experimental data from Inyushkin \emph{et al.} \cite{Inyushkin2004}, with fitting parameters shown in Table \ref{table:parameters}. Kinetic and collective thermal conductivity regimes for $^{\rm na}$Si are also plotted in dashed lines.}
\label{fig:natural_bulk_conductivity}
\end{figure}

In the following, we compare our predictions with experimental data on silicon samples of different sizes at a large temperature range. For calculations we use Eq. (\ref{kappa_final}) where (\ref{conductivity_kin_dos}) and (\ref{conductivity_coll_dos}) are the corresponding kinetic and collective terms, (\ref{sigma}) is used for the switching factor and (\ref{factor_F}) for the form factor. In all these expressions we use expressions (\ref{tauI})-(\ref{tauU}) for the relaxation times. Group velocities are always calculated from dispersion relations.
Results using Eq. (\ref{kappa_final}) and (\ref{tauI})-(\ref{tauU}) are plotted in Fig. \ref{fig:natural_bulk_conductivity} and compared to data from the work by Inyushkin \textit{et al.}\cite{Inyushkin2004} for natural and enriched 99.983$\%$ $^{28}$Si ($^{\rm na}$Si and $^{\rm iso}$Si). The parameters $B_N, B_{N}$ and $B_U$ appearing in the phonon-phonon relaxation times (\ref{tauN})-(\ref{tauU}) are obtained by fitting  $^{\rm na}$Si sample and their values are shown in Table \ref{table:parameters}. The same values are  used for the enriched sample. The remaining scattering rates (\ref{tauI})-(\ref{tauB}) are free of adjustable parameters. Both samples ($^{\rm na}$Si and $^{\rm iso}$Si) are reported to have the same effective size  $L_{\rm eff}=2.8$ mm. We have used a mass-fluctuation factor of $\Gamma_{^{\rm na}\rm Si }=2.01\times10^{-4}$ and  $\Gamma_{^{\rm iso}\rm Si}=\Gamma_{^{\rm na}\rm Si}/625=3.2\times10^{-7}$ respectively. Note that the position of the peak for both $^{\rm na}$Si is correctly fitted and for $^{\rm iso}$Si is correctly predicted (solid lines overlap experimental points in the plot) being the only change between both samples the calculated mass-fluctuation factor. This is a proof of the consistency of our model and confirms the prediction given by  Inyushkin \textit{et al}.\cite{Inyushkin2004}. In the following subsection we have done the same test for Callaway and pure RTA models, obtaining worse results  (see Fig. \ref{fig:comparison}).

In Fig. \ref{fig:natural_bulk_conductivity} we also show the limiting curves corresponding to kinetic $\kappa_{\rm kin}$ and collective regime $\kappa_{\rm coll}$ for $^{\rm na}$Si according to Eqs. (\ref{conductivity_kin_dos}) and (\ref{conductivity_coll_dos}) respectively.  It can be seen that in the low temperature range the sample is entirely in the kinetic regime, since boundary is expected to dominate over normal scattering. $\kappa$ tends to the collective regime as temperature rises and N-processes begin to be dominant.  In the collective regime all the phonons notice the scattering events suffered by the rest of the collectivity, thus the thermal conductivity is significantly lower than in the kinetic regime. At this point one can notice the first important implication of the present formulation. Both limits contain only resistive terms in their integrals, but $\kappa_{\rm coll}$ is less conductive than $\kappa_{\rm kin}$. This seems to be in contradiction with the fact that $\kappa_{\rm coll}$ is governed by normal scatterings and this has a non-resistive nature. Actually the ability of N-processes at distributing the energy between modes enhances the resistive character of the rest of the scattering mechanisms. This physics can be understood thanks to the different mathematical treatment of the relaxation times inside the integrals, interpreted in terms of serial and parallel resistivities in Sec. \ref{sec:thermal_conduct}. Our model allows to understand this unlike Callaway model where normal scattering is considered inside the resistive integrals.
Another remarkable behavior is the dominance of normal scattering even at room temperature. One can expect umklapp processes to dominate at high temperatures, but it can be seen that is not the case of bulk silicon at room temperature. It can be observed in Fig. \ref{fig:natural_bulk_conductivity} the curves seem to suggest a change in the tendency at high temperature regime. $\kappa$ seems to tend to a more kinetic behavior at very high temperature. The temperature range where kinetic regime happens at high temperature will depend on the height of the dispersion relations that eventually determines the importance of umklapp respect to normal scattering.

\begin{figure}[htb]
\includegraphics[width=0.45\textwidth]{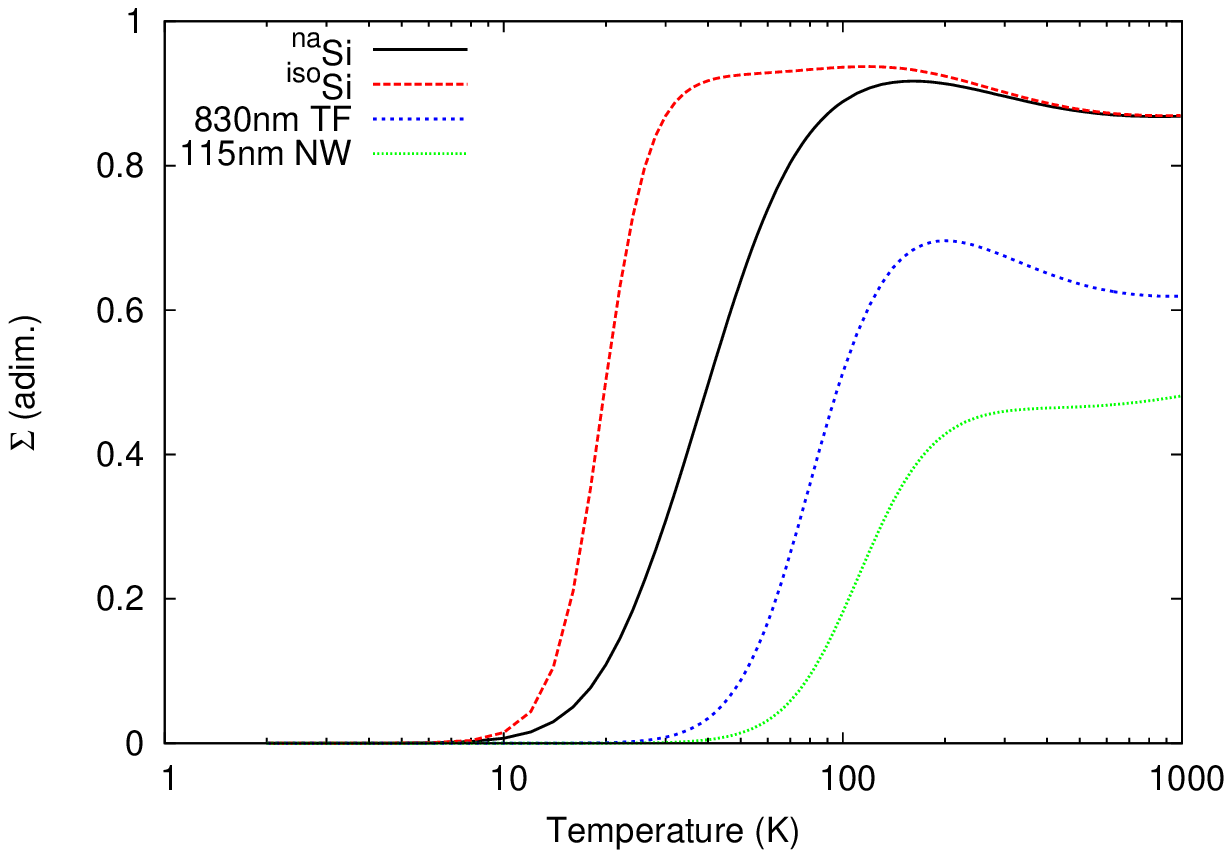}
\caption{(Color online) Switching factor $\Sigma$ as a function of temperature in a semilogarithmic plot for $^{\rm na}$Si and $^{\rm iso}$Si bulk, 830nm TF and 115nm NW.}
\label{fig:sigma}
\end{figure}

The transition from one regime to another is determined by $\Sigma$, shown in Fig. \ref{fig:sigma}. At very low temperatures, the boundary scattering present in $\tau_{R}$ behaves as $\tau_{B}\sim L_{\rm eff}\ll\tau_{N}$ and yields $\Sigma=0$, we are clearly in the kinetic regime $\kappa\sim \kappa_{\rm kin}$. At room temperature we can easily calculate the ratio of $\tau_N/\tau_U$ (neglecting all other scattering mechanisms) and realize that it is of the order of 0.1, thus $\Sigma\approx 1$ (actually $\Sigma= 0.9$) and we are in the collective regime, $\kappa\sim \kappa_{\rm coll}$.  It can be observed in Fig. \ref{fig:sigma} that for $^{\rm iso}$Si the transition to the collective regime is sharper than for $^{\rm na}$Si. This is due to the fact that, for these samples, the transition happens in the region of impurity scattering dominance.

\begin{figure}[htb]
\includegraphics[width=0.45\textwidth]{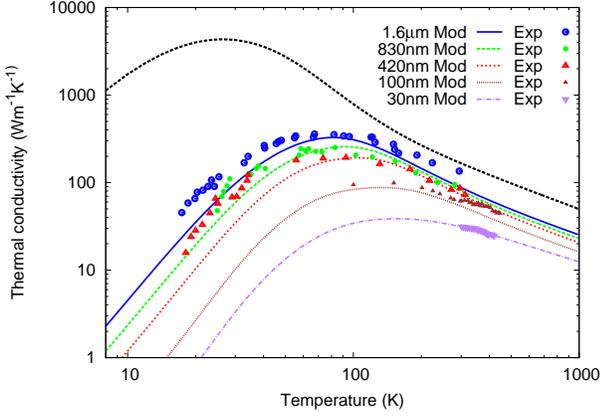}
\caption{(Color online) Thermal conductivity of different thicknesses (1.6 $\mu$m, 830 nm, 420 nm, 100 nm  and 30 nm) Si thin films as a function of temperature in a double logarithmic plot. Model predictions are shown in lines according to the legend. Experimental data\cite{Asheghi1998} are shown in symbols also according to the legend. $^{na}$Si bulk thermal conductivity is plotted for reference (black dashed line).}
\label{fig:conductivityTF}
\end{figure}

\begin{figure}[htb]
\includegraphics[width=0.45\textwidth]{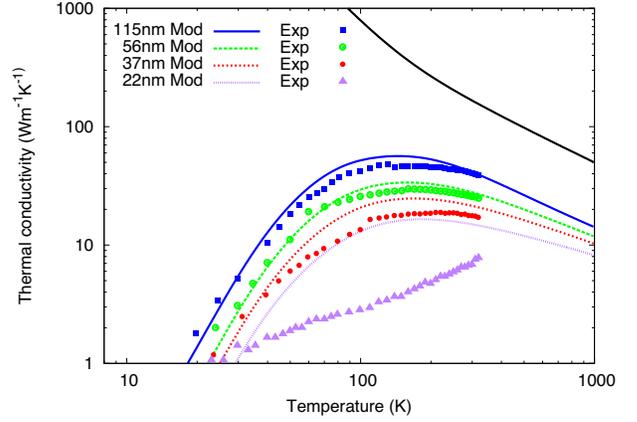}
\caption{(Color online) Thermal conductivity of different diameter Si nanowires as a function of temperature in a double logarithmic plot. Model predictions for different diameters (115 nm, 56 nm, 37 nm and 22 nm) are shown in lines according to the legend. Experimental data\cite{Li2003} are shown in symbols also according to the legend. $^{na}$Si bulk thermal conductivity is plotted for reference (black solid line).}
\label{fig:conductivityNW}
\end{figure}

Results for silicon TFs and NWs are shown in Figs. \ref{fig:conductivityTF} and \ref{fig:conductivityNW} respectively. TFs are those from the work by Asheghi \textit{et al.}\cite{Asheghi1998}, with thicknesses of $h=1.6 \mu$m, 830 nm, 420 nm, 100 nm  and 30 nm.  Their respective effective lengths are thus $L_{\rm eff}=3.6 \mu$m, 1.87 $\mu$m, 945 nm, 225 nm and 67.5 nm. NWs are those from the work by Li \textit{et al.}\cite{Li2003} with diameters $d=$ 115 nm, 56 nm, 37 nm and 22nm. In this case $L_{\rm eff}$ is equivalent to the diameters. The rest of the parameters remain the same as in the case of $^{\rm na}$Si bulk.  It can be observed that all curves are in good agreement with the experimental data with the exception of the thinnest NW (22 nm) and in some intermediate temperature region for the 37 nm NW.  Note that all these samples may contain a certain concentration of impurities due to fabrication process\cite{Asheghi1998}, but we have maintained the mass-fluctuation factor $\Gamma_{^{\rm na}\rm Si}$ for all the nanoscale samples because there is no reported data about this question. From the plots we can confirm that Eq. (\ref{kappa_final}) is able to correctly describe thermal conductivity behavior for general geometries and sizes without the inclusion of confinement effects above an effective size of 30 nm.

\begin{figure}[htb]
\includegraphics[width=0.45\textwidth]{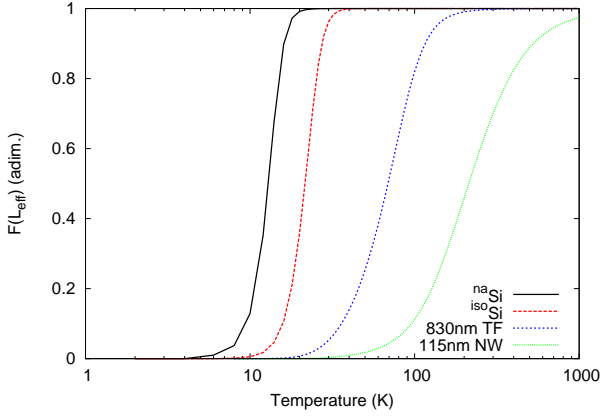}
\caption{(Color online) Geometric factor $F(L_{\rm eff})$ as a function of temperature in a semilogarithmic plot for $^{\rm na}$Si and $^{\rm iso}$Si bulk, 830nm TF and 115nm NW.}
\label{fig:factorF}
\end{figure}

Furthermore, it is clear from Fig. \ref{fig:sigma} that the smaller $L_{\rm eff}$ the more kinetic $\kappa$ is. This is reasonable and expected, since at reduced sizes boundary scattering rate should contribute the most to thermal resistance not only at low temperature but also at room temperature. The size effects are illustrated through the form factor $F(L_{\rm eff})$ plotted in Fig. \ref{fig:factorF}.

\begin{table}[b]
\caption{Fitting parameters for $^{\rm na}$Si bulk}
\centering
\begin{tabular}{c c c c}
\hline\hline
& $B_U$ (s$^3$K$^{-1}$)  & $B_{N}$ (sK$^{-3}$) & $B'_{N}$ (s$^{-1}$K$^{-1}$)  \\ [0.5ex]
\hline
This model & $2.8\times 10^{-46}$ &  $3.9\times 10^{-23}$ & $4.0\times 10^8$  \\ [0.5ex] 
\hline 
Callaway model & $1.4\times 10^{-46}$ &  $3.5\times 10^{-24}$ & $1.0\times 10^7$  \\ [0.5ex]
standard RTA & $1.9\times 10^{-45}$ &  $9.3\times 10^{-23}$ & $3.2\times 10^5$  \\ [0.5ex]
\hline
\end{tabular}
\label{table:parameters}
\end{table} 

It can be noticed from the plots that if one hopes to fit the experimental values with a pure kinetic expression, extra thermal resistivity should be added in room and high temperature regions to reduce the predicted values.  The  presence of the collective term in our Eq. (\ref{kappa_final}) makes unnecessary this adjustment. On the contrary, our model explains why VM models should give poor results at low temperatures. The boundary term leads the system to a kinetic regime at these temperatures, raising the thermal conductivity.

Obviously, our phenomenological expressions for the relaxation times cannot be used to obtain an extremely accurate fit. Further improvements of the model can be achieved by a more precise treatment of scattering times through \textit{ab initio} techniques, but we have demonstrated that some issues related to relaxation times come from their incorrect averaging.  We can conclude that an appropriate treatment of the N-processes makes unnecessary the introduction of new terms in the expression of $\kappa$. Probably rough surfaces\cite{Hochbaum2008} would need additional considerations to improve the fit but this is out of the scope of the present work.

\subsection{Comparison with other models}
\begin{figure}[htb]
\includegraphics[width=0.45\textwidth]{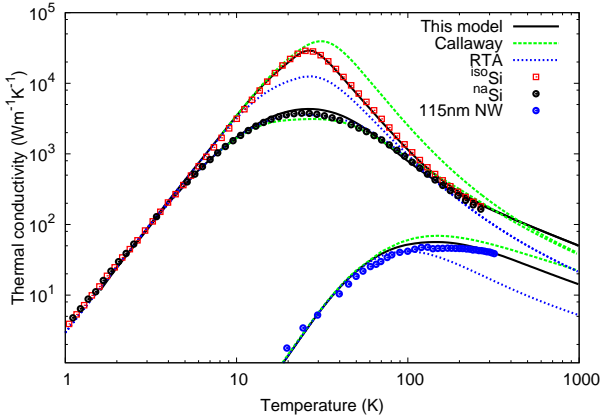}
\caption{(Color online) Thermal conductivity $\kappa$ as a function of temperature for $^{\rm na}$Si and $^{\rm iso}$Si bulk, and 115nm NW. Best fit provided by standard RTA and Callaway models against this work model.}
\label{fig:comparison}
\end{figure}

In order to show the improvement of our model over standard RTA and Callaway model\cite{Callaway1958}, we compare our results with those obtained with these usual approaches. The procedure we have followed to fit $^{\rm na}$Si is the same as in our approach. The same relaxation times equations (\ref{tauB})-(\ref{tauU}) are used in the three approaches to highlight only the models accuracy. The values of the fitting parameters that provide the best results for $^{\rm na}$Si in each approach are shown in Table \ref{table:parameters}. Then, to test the prediction capability, we have changed the mass-fluctuation factor for ($^{\rm iso}$Si and the effective size for the 115nm NW.  Results are provided in Fig. \ref{fig:comparison}.

As expected, RTA reproduces very well  $^{\rm na}$Si in the low temperature range, but from $T>200K$ begins to diverge from experimental data. However it  underpredicts  the $^{\rm iso}$Si peak and from this point forward. At the nanoscale it also fails in the prediction as shown in the plot. On the other hand, although Callaway model is able to reproduce correctly $^{\rm na}$Si sample, it overpredicts $^{\rm iso}$Si and the 115 nm nanowire. 

In the literature we can find two kind of approaches. Firstly we have models that focus on the fitting to natural and isotopically enriched bulks in the whole or partial temperature range, but they are  not proved at the nanoscale.\cite{Morelli2002} On the opposite way, we can find models focused on the fitting to the nanoscale giving a good agreement with measurements but they are not proved at reproducing other isotopic composition bulks.\cite{Mingo2003,Chantrenne2005} Since providing a good fit at the peak region for both bulks is very difficult, most of the published models do not show the corresponding temperature interval. Normally they show fits and predictions from $T>50K$. With 4 simple and representative scattering events (boundary, impurities, normal and umklapp) our model is able to provide a very satisfactory fit from the macro to the  
nanoscale in the whole range of temperatures.
\begin{figure}[htb]
\includegraphics[width=\columnwidth]{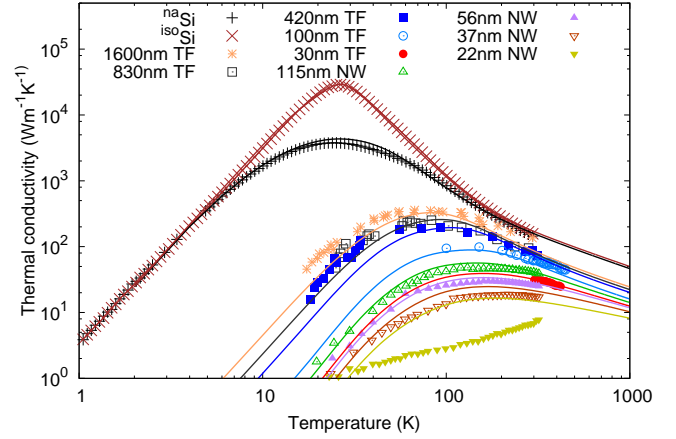}
\caption{(Color online) Thermal conductivity of all the silicon samples studied in this work (Bulk, thin films and nanowires). It can be seen that a very good global agreement is obtained at all ranges of size and temperature.}
\label{fig:conductivity_tot}
\end{figure}
 In Fig. \ref{fig:conductivity_tot} we show the global prediction achieved by our model, with this plot one can notice in a single view how the thermal conductivity works for the complete set of different size, shape and composition Si samples in the [1-1000]K temperature interval.

\section{Conclusions}

This work shows that the key point for an accurate description of the thermal conductivity in the whole range of temperatures is taking into account the effect of normal processes on the phonon collective behavior. As a consequence two well differentiated thermal transport regimes are studied for the first time, kinetic and collective, depending on the relative importance of normal processes.

The proposed model gives an expression of $\kappa$ valid for all ranges of temperatures. This expression is obtained by combining the VM and RTA approaches including a switching factor that determines the transport regime in terms of the normal and resistive mean scattering times. In these regimes, differences in the phonon averaging and in the way to account for the boundary effects are considered.

We have also included higher-order non-equilibrium effects through an analytical function $F(L_{\rm eff})$ to generalize the model to any kind of sample depending on its geometry and characteristic size. The obtained results agree very well with experimental measurements of different Si samples of characteristic length above 30 nm, proving that above this size quantum confinement effects are not necessary to explain thermal transport.

\acknowledgements
The authors acknowledge financial support from projects CSD2010-00044, FIS2012-32099, MAT2012-33483, and 2009-SGR00164, and from a Marie Curie Reintegration Grant. The authors thank Profs. D. Jou, J. Camacho and J. Bafaluy for fruitful discussions and M. M. de Lima Jr for a critical reading of the manuscript. Thanks are also given to the Red Espa\~{n}ola de Supercomputaci\'{o}n providing access to the supercomputer TIRANT.


\end{document}